\documentclass[prb,aps,twocolumn,showpacs,superscriptaddress,longbibliography]{revtex4-2}
\usepackage{graphicx}
\usepackage{graphicx, color}
\usepackage{hyperref}
\definecolor{link}{rgb}{0.1,0.1,0.9}
\hypersetup{colorlinks=true,linkcolor=link,citecolor=link,urlcolor=link,linktocpage}
\usepackage{epsfig}
\usepackage{amsmath}
\usepackage{amsfonts}
\usepackage{amssymb}
\usepackage{graphicx}
\usepackage{color}
\usepackage{tabularx}
\usepackage{txfonts}
\usepackage{bm}
\usepackage{dcolumn}  

\begin{document}
\title{N\'eel order, spin-spiral, and spin liquid ground state in frustrated three dimensional system CaMn$_2$P$_2$: A DFT+U and spin dynamics study}
\author{Bidyut Mallick}
\thanks{These authors contributed equally to this work.}
\affiliation{Department of Applied Sciences, Galgotias College of Engineering and Technology, Greater Noida, India}
\author{Sk. Soyeb Ali}
\thanks{These authors contributed equally to this work.}
\affiliation{Department of Physics, Bennett University, Greater Noida-201310, Uttar Pradesh, India.}
\author{S. K. Panda}
\email{Electronic address: swarup.panda@bennett.edu.in}
\affiliation{Department of Physics, Bennett University, Greater Noida-201310, Uttar Pradesh, India.}

\begin{abstract}
We investigate the magnetic ground state and phase transitions in the frustrated three-dimensional system CaMn$_2$P$_2$ using first-principles calculations combined with spin-dynamics simulations. Our density functional theory (DFT) calculations, incorporating Hubbard $U$ corrections, reveal that CaMn$_2$P$_2$ exhibits an indirect gap semiconducting ground state with a localized Mn$^{2+}$ ($3d^5$, $S=\frac{5}{2}$) electronic configuration and negligible spin-orbit coupling effects. The computed exchange interactions show that the magnetic behavior is well described by a isotropic Heisenberg Hamiltonian. In this model, there are two major couplings: the nearest-neighbor (NN) interaction $J_1$ couples the two Mn layers along the $c$-axis and next NN $J_2$ is in the $a$-$b$ plane where Mn ions form a hexagonal layer structure. Our results show that both $J_1$ and $J_2$ are antiferromagnetic in nature and as a consequence $J_2$ induce frustration owing to the in-plane triangular geometry of the Mn-ions. The $J_1$ is found to promote long-range antiferromagnetic order, while the $J_2$ is responsible for spin canting and disorder. Our spin-wave analysis confirms that the system stabilizes a spin-spiral ground state with a propagation vector $q = (\frac{1}{6}, \frac{1}{6}, 0)$ in agreement with neutron diffraction experiments. By tuning the $\frac{J_2}{J_1}$ ratio, we construct a phase diagram that reveals a transition from a collinear N\'eel antiferromagnetic state to spin-spiral phases with different propagation vectors, and eventually to a disordered phase at large frustration. Atomistic spin-dynamics simulations capture the temperature evolution of the magnetism and reproduce the experimentally measured magnetic specific heat as well as transition temperature with good accuracy. Furthermore, for large $\frac{J_2}{J_1}$, we identify a low-temperature phase with slow spin relaxation and persistent fluctuations, suggesting a spin-liquid-like state. Our study provides a microscopic understanding of frustration-induced magnetism in CaMn$_2$P$_2$ and establishes it as a realization of $J_1$-$J_2$ model in three-dimensional lattice for exploring emergent magnetic phases.
\end{abstract}
\maketitle

\section{Introduction}
Magnetic structure with frustrated spin state have attracted lot of interest because of their potential applications in emerging technology and fundamental perspective~\cite{RevModPhys.82.53, GOBEL20211,doi:10.1021/acs.chemrev.0c00297}. Competing magnetic interactions at various spatial ranges, dictated by crystal geometry, induce magnetic frustration. A representative example of spin frustration from competing interactions is bipartite honeycomb lattice. In these types of lattices, competition between nearest-neighbour interaction (J$_1$) and next-nearest-neighbour interactions (J$_2$) leads to variety of complex magnetic states ~\cite{RASTELLI19791, Katsura1986}. In the classical limit, N\'eel antiferromagnet, stripy, zigzag and spiral magnetic order have been reported for honeycomb lattice ~\cite{PhysRevB.81.214419,PhysRevB.84.094424, Bishop_2012}. Further, recent experimental study on the bilayer honeycomb-lattice reported spin-liquid-like behavior in the low-temperature region~\cite{doi:10.1021/ja901922p,PhysRevLett.105.187201, Okumura2010}. 
\par
\begin{figure}[t]
\includegraphics[width=1\columnwidth]{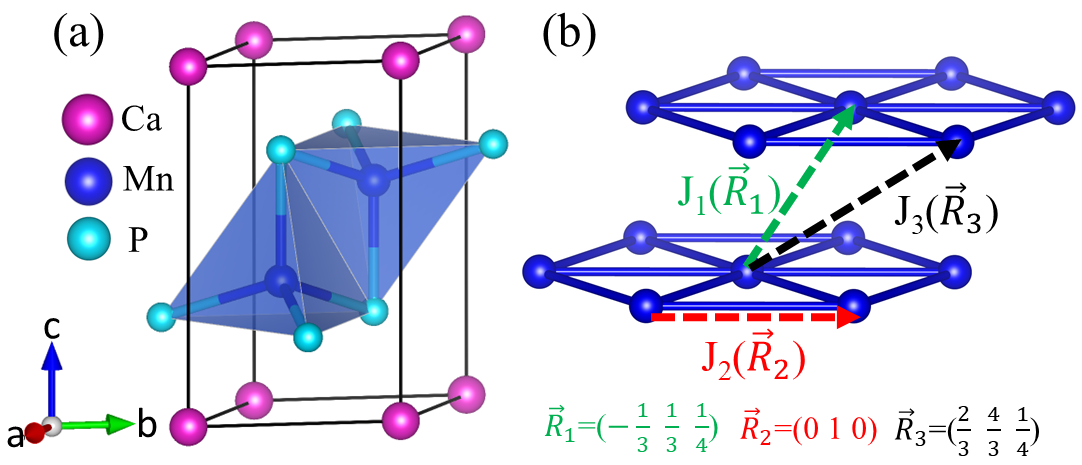}
\caption{ (a) The the unit cell of CaMn$_2$P$_2$ where two Mn ions are connected via edge shared MnP$_4$ tetrahedral units, (b)
A schematic representation of the triangular Mn network in the $a$-$b$ plane and their stacking along the $c$-direction. The Mn-Mn exchange interactions corresponding to various neighbors are also marked.} 
\label{fig1}
\end{figure}

In this context, the Mn-based 122-type bipartite honeycomb systems have received lot of attention in the recent times due to their diverse magnetic properties. BaMn$_2$Pn$_2$ (Pn = P, As, Sb, Bi) compounds, crystallize in the tetragonal ThCr$_2$Si$_2$ type structure (I4/mmm), are reported to undergo G-type AFM ordering~\cite{BROCK1994303,PhysRevB.79.094519,SAPAROV201332,PhysRevB.80.100403}. On the other hand, CaMn$_2$Pn$_2$ crystallizes in the trigonal CaAl$_2$Si$_2$-type structure (see Fig. ~\ref{fig1}(a)) with space group P3$\overline{m}$1 ~\cite{doi:10.1073/pnas.2108724118,PhysRevB.94.094417,PhysRevB.86.184430,Bridges2009,PhysRevB.91.085128}. In CaAl$_2$Si$_2$-type structure, Mn sublattice form a two-dimensional triangular lattice in the $a$-$b$ plane (Fig.~\ref{fig1}(b)), while Pn atoms occupy the interstitial positions, coordinating with Mn to form edge-shared MnP$_4$ tetrahedra (Fig.~\ref{fig1}(a)). This is inherently prone to magnetic frustration provided that the Mn-Mn couplings within the triangular network are antiferromagnetic in nature. The alternating layers of Ca atoms and Mn$_2$Pn$_2$ units stacked along the $c$-axis (Fig.~\ref{fig1}). This arrangement results in a highly anisotropic structure, where the in-plane interactions among Mn atoms compete with the interlayer couplings. The presence of competing magnetic couplings often lead to unconventional magnetic ground states, including spin-spiral states. This is evidenced form the recent neutron diffraction study that revealed CaMn$_2$As$_2$, CaMn$_2$Bi$_2$ and CaMn$_2$Sb$_2$ as N\'eel-antiferromagnet~\cite{PhysRevB.91.085128, Bridges2009,PhysRevB.91.180407}, whereas CaMn$_2$P$_2$ has been reported to show spin-spiral ground state ~\cite{PhysRevB.107.054425}. CaMn$_2$P$_2$ exhibits first-order antiferromagnetic (AFM) transition at T$_N$ = 70 K implied by specific heat C$_p$(T), magnetic susceptibility $\chi$(T) and NMR measurements ~\cite{doi:10.1073/pnas.2108724118,Li_2020,PhysRevB.107.054425}, whereas the isostructural CaMn$_2$As$_2$, CaMn$_2$Sb$_2$ and CaMn$_2$Bi$_2$ compounds undergo second-order AFM transitions ~\cite{PhysRevB.94.094417,PhysRevB.86.184430,PhysRevB.91.180407,PhysRevB.91.085128}. The neutron-diffraction study considered a three-state Potts-nematic order with complex spiral magnetic state that breaks threefold rotational symmetry in CaMn$_2$P$_2$ below the transition temperature to explain first order phase transition ~\cite{PhysRevB.107.054425}. Another study by Sangeetha \textit{et. al.}~\cite{doi:10.1073/pnas.2108724118} suggested that strong first order transition may originate from the magnetic frustration due to bipartite Mn layers. An alternative explanation is proposed by optical measurement, where authors suggested a weak partial re-normalization of band structure with the occurrence of van Hove singularity caused by the phase transition~\cite{Zheng2023}. Raman spectrum of CaMn$_2$P$_2$ measured at T = 10 K showed new peaks in low frequency range along with the peaks detected at high temperature. However single crystal X-ray diffraction showed no difference in the crystal structure at 293 and 40 K. This was explained by considering formation of superstructure below the transition temperature~\cite{Li_2020}. The formation of superstructure have been discarded by high-resolution single-crystal synchrotron X-ray diffraction studies at 20 K ~\cite{doi:10.1073/pnas.2108724118}. Thus, the physical origin of the strong first order transition in CaMn$_2$P$_2$ still remain in debate. Further the recent Neutron diffraction measurements~\cite{PhysRevB.107.054425} reveal a spin-spiral ground state with propagation vector $q$ = ($\frac{1}{6}, \frac{1}{6}, 0$), corresponding to an in-plane spiral with a six-site periodicity along both $a$ and $b$ directions. Within each triangular Mn layer, the moments rotate smoothly in-plane, forming a 360$^\circ$ spiral over six Mn sites. Adjacent layers within each Mn bilayer are antiferromagnetically coupled, aligning their spirals in an antiparallel fashion. The electrical resistivity measurement, and optical study~\cite{Zheng2023} revealed that CaMn$_2$P$_2$ has semi-conducting ground state. 
\par 
Despite the experimental advancement in understanding the complex magnetism of CaMn$_2$P$_2$, there is currently a lack of comprehensive first-principles investigations that explore its magnetic ground state, electronic structure, and the underlying exchange interactions. These properties are critical for understanding its complex magnetic ground state. The interesting structural framework of CaMn$_2$P$_2$ also allows for strong interplay between the crystal lattice, electronic states, and magnetic properties, making it an intriguing material for exploring unconventional magnetic behavior. Thus, we report a detail analyses based on density functional theory calculations including Hubbard $U$ corrections (DFT+U approach) and atomistic spin-dynamics simulations, that provide a reliable framework for uncovering the magnetic ground state of a system. Our analysis will not only elucidate the magnetic and electronic properties of CaMn$_2$P$_2$ but also contribute to the broader understanding of frustrated three-dimensional magnetic systems with competing interactions. We demonstrate that interplay between crystal structure, electronic correlations, and inter-atomic magnetic couplings lead to a complex phase diagram with various possible exotic magnetic ground states, advancing our knowledge about magnetically frustrated systems. 

\section{Methods}
The calculations reported in this work are carried out using density functional theory (DFT) based approaches~\cite{DFT1,DFT2}. We employed two complementary full-potential methods, namely the full-potential linearized augmented plane wave (FP-LAPW) method, as implemented in the WIEN2K code \cite{SCHWARZ2003259}, and the full-potential linearized muffin-tin orbital (FP-LMTO) method, as implemented in the RSPt code \cite{PhysRevB.12.3060, PhysRevB.36.3809, wills2000electronic}. The exchange and correlation effects are treated within generalized gradient approximation (GGA)~\cite{GGA_PBE} and an orbital dependent Hubbard $U$ term is added for the strongly correlated Mn-$d$ orbitals in the framework of  GGA+U approach~\cite{GGA+U}. Based on the suggestion of earlier studies~\cite{PhysRevMaterials.5.104402,PhysRevB.87.165118}, we have considered $U$ = 4 eV and Hund's $J$ = 0.8 eV for the Mn$-d$ states. The Brillouin-zone integration was done with a modified tetrahedron method~\cite{tetrahedron} using a $10\times10\times5$ $k$-mesh to achieve self-consistency. The consistency of the results from these two methods provides further credence to our calculations. The effect of spin-orbit coupling (SOC) for valence states was also included in an accurate second-variational method using scalar relativistic wave functions.  Since SOC effects are found negligible, reported results are from GGA+U simulations. 
\par 
 After obtaining good convergence of charge density in our FP-LMTO calculations within GGA+U approach, The effective inter-site exchange parameters $J_{ij}$ between Mn spins at $i^{th}$ and $j^{th}$ site were obtained by employing the magnetic force theorem\cite{magnetic-force1,magnetic-force2}. In this method, the total converged energies of the magnetic system was mapped onto the following Heisenberg spin Hamiltonian
\begin{equation}
\ H =  -\frac{1}{2}\sum_{i\neq j}J_{ij}(\hat{S}_i \cdot \hat{S}_j) 
\end{equation}
where $\hat {S}_i$, $\hat {S}_j$ are spin unit vectors for the ${i}^{th}$ and ${j}^{th}$ site, respectively. Finally, the effective $J_{ij}$s are extracted in a linear-response manner via a Green’s function technique using the following formula: 
\begin{equation}
J_{i j}=\frac{\mathrm{T}}{4} \sum_n \operatorname{Tr}\left[\hat{\Delta}_i\left(i \omega_n\right) \hat{G}_{i j}^{\uparrow}\left(i \omega_n\right) \hat{\Delta}_j\left(i \omega_n\right) \hat{G}_{j i}^{\downarrow}\left(i \omega_n\right)\right],
\end{equation}
where $\Delta_i$ is the onsite spin splitting, the spin-dependent intersite Green’s function is $G_{ij}$, and $\omega_n$ are the $n^{th}$ Fermionic Matsubara frequencies. A detailed discussion of the implementation of the magnetic force theorem in RSPt is provided in Ref.~\onlinecite{PhysRevB.91.125133}. This method has been successfully used for many other transition metal compounds~\cite{PhysRevB.94.064427,PhysRevB.109.035125,PhysRevB.106.L180408}.
\par
We constructed the spin Hamiltonian using the computed inter-site exchange parameters, $J_{ij}$ and determined the magnetic ground state via linear spin wave theory, as implemented in the SpinW code \cite{Toth_2015}. To investigate finite-temperature effects and dynamical evolution of spin configurations, we performed atomistic spin dynamics (ASD) simulations based on the stochastic Landau-Lifshitz-Gilbert (sLLG) formalism as implemented in UppASD code~\cite{PhysRevB.54.1019, Skubic_2008}. The magnetization dynamics of each atomic moment $\vec{m}_i$ are governed by the following sLLG equation:
\begin{equation}
\frac{\partial \vec{m}_i}{\partial t} = -\frac{\gamma}{1 + \alpha^2} \left( \vec{m}_i \times \vec{B}^{\mathrm{eff}}_i + \frac{\alpha}{m} \vec{m}_i \times \left( \vec{m}_i \times \vec{B}^{\mathrm{eff}}_i \right) \right),
\end{equation}
where $\gamma$ is the gyromagnetic ratio, $\alpha$ is the dimensionless Gilbert damping parameter. The total \textit{effective magnetic field} $\vec{B}^{\mathrm{eff}}_i$ acting on site $i$ consists of a deterministic part $\vec{B}_i$, derived from the system’s spin Hamiltonian $H$, and a stochastic term $\vec{B}^{\mathrm{fl}}_i(t)$, which models thermal fluctuations:
\[
\vec{B}^{\mathrm{eff}}_i = \vec{B}_i + \vec{B}^{\mathrm{fl}}_i(t), \quad \text{with} \quad \vec{B}_i = -\frac{\partial H}{\partial \vec{m}_i}.
\]
The stochastic magnetic field $\vec{B}^{\mathrm{fl}}_i(t)$ is treated as a Gaussian white noise with the following statistical properties:
\[
\langle B^{\mathrm{fl}}_i(t) \rangle = 0, \quad \langle B^{\mathrm{fl},k}_i(t) B^{\mathrm{fl},p}_j(t') \rangle = 2D\, \delta_{ij} \delta_{kp} \delta(t - t'),
\]
where $i, j$ label lattice sites, $k, p$ denote Cartesian components, and $D$ characterizes the strength of thermal fluctuations:
\[
D = \frac{\alpha}{1 + \alpha^2} \frac{k_B T}{\mu_B m}.
\]
This thermal noise term ensures that the system evolves according to the fluctuation–dissipation theorem and properly samples thermally activated spin configurations. For all simulations, we set $\alpha = 1.0$ to accelerate convergence toward equilibrium. The simulations are performed using a time step of $\Delta t = 0.1$ femtoseconds, providing adequate resolution for capturing both fast and slow spin dynamics. The system was initially prepared in a high-temperature disordered spin configuration at T = 300 K, representing a fully randomized state. It was then gradually cooled to 0 K in uniform temperature decrements of 2 K, closely mimicking a quasi-adiabatic annealing process. At each temperature step, the spin dynamics were evolved over several thousand integration steps, allowing the system to locally equilibrate under the influence of both deterministic precessional dynamics and stochastic thermal fluctuations. The evolution at each temperature step was governed by the stochastic LLG equation. The final spin configuration obtained after equilibration at a given temperature was used as the initial condition for the next lower temperature step. This sequential propagation of spin states across decreasing temperatures enables the system to explore the energy landscape and gradually relax into lower-energy configurations. The use of small temperature steps and long integration times at each step ensures that the system evolves toward its equilibrium magnetic state in a manner that approximates adiabatic cooling, avoiding artificial trapping in metastable configurations. This approach enabled us to explore the equilibrium spin configurations, magnetic phase transitions, and relaxation dynamics.
\par 
The structure parameters used in our calculations were taken from the experimental work as reported in Ref.~\cite{doi:10.1073/pnas.2108724118}. CaMn$_2$P$_2$ crystallizes in the trigonal structure (spacegroup no. 164)) as shown in Fig.~\ref{fig1}(a). Each Mn atom is coordinated by four phosphorus atoms in a distorted tetrahedral environment. The magnetic Mn atoms occupy two positions per unit cell, forming two adjacent layers of triangular lattices in the $a$-$b$ plane as shown in Fig.~\ref{fig1}(b). These layers are stacked along the $c$-direction, forming closely spaced magnetic bilayers. Each bilayer is separated from the next by nonmagnetic Ca–P slabs, resulting in a large interlayer separation of approximately 5.67 $\AA$ between adjacent bilayers. This structural motif gives rise to a  magnetic lattice, in which the dominant exchange interactions are largely confined within each magnetic double layer. The magnetic interactions are described by three key exchange parameters, labeled as $J_1$, $J_2$ and $J_3$ in Fig.~\ref{fig1}(b), with the corresponding Mn–Mn displacement vectors $\overrightarrow{R_1}$, $\overrightarrow{R_2}$ and $\overrightarrow{R_3}$, respectively. The first-nearest-neighbor interaction $J_1$ connects Mn atoms across the two adjacent layers of the bilayer via the vector $\overrightarrow{R_1} = (\frac{-1}{3}, \frac{1}{3}, \frac{1}{4})$ with a Mn–Mn separation of 2.92 \AA. The second-nearest-neighbor interaction, $J_2$, couples Mn atoms within the same triangular layer in the $a$-$b$ plane, through the vector $\overrightarrow{R_2} = (0, 1.0, 0)$, with a separation of 4.10 \AA. The third-nearest-neighbor interaction, $J_3$ connects Mn atoms in the bilayer through the vector $\overrightarrow{R_3} = (\frac{2}{3}, \frac{4}{3}, \frac{1}{4})$, spanning a distance of 5.03 \AA.

\section{Results and Discussion}
\noindent\textbf{Electronic structure and magnetism using first-principles calculations:}
We began our analysis by calculating the total energies for two distinct magnetic configurations: the ferromagnetic state, in which the Mn spins within the unit cell are aligned parallel to each other, and the antiferromagnetic state, where the Mn spins are oriented antiparallel to each other. Our results as obtained from the GGA+U approach find that the antiferromagnetic state is lower in energy by 303 meV per Mn and the spin moment on Mn is 4.28 $\mu_B$. Such a high value of  of moment emphasizes the localized nature of the Mn-$d$ orbitals. We also included the effect of SOC and computed the orbital moment of Mn ions which comes out to be negligible (0.001$\mu_B$). The full quenching of the orbital moment is expected since $d$-orbitals of Mn is half-filled in CaMn$_2$P$_2$. Thus, we conclude that the SOC does not play any role and rest of the analyses are carried out using GGA+U approach. 
\par
\begin{figure}[t]
\includegraphics[width=1.0\columnwidth]{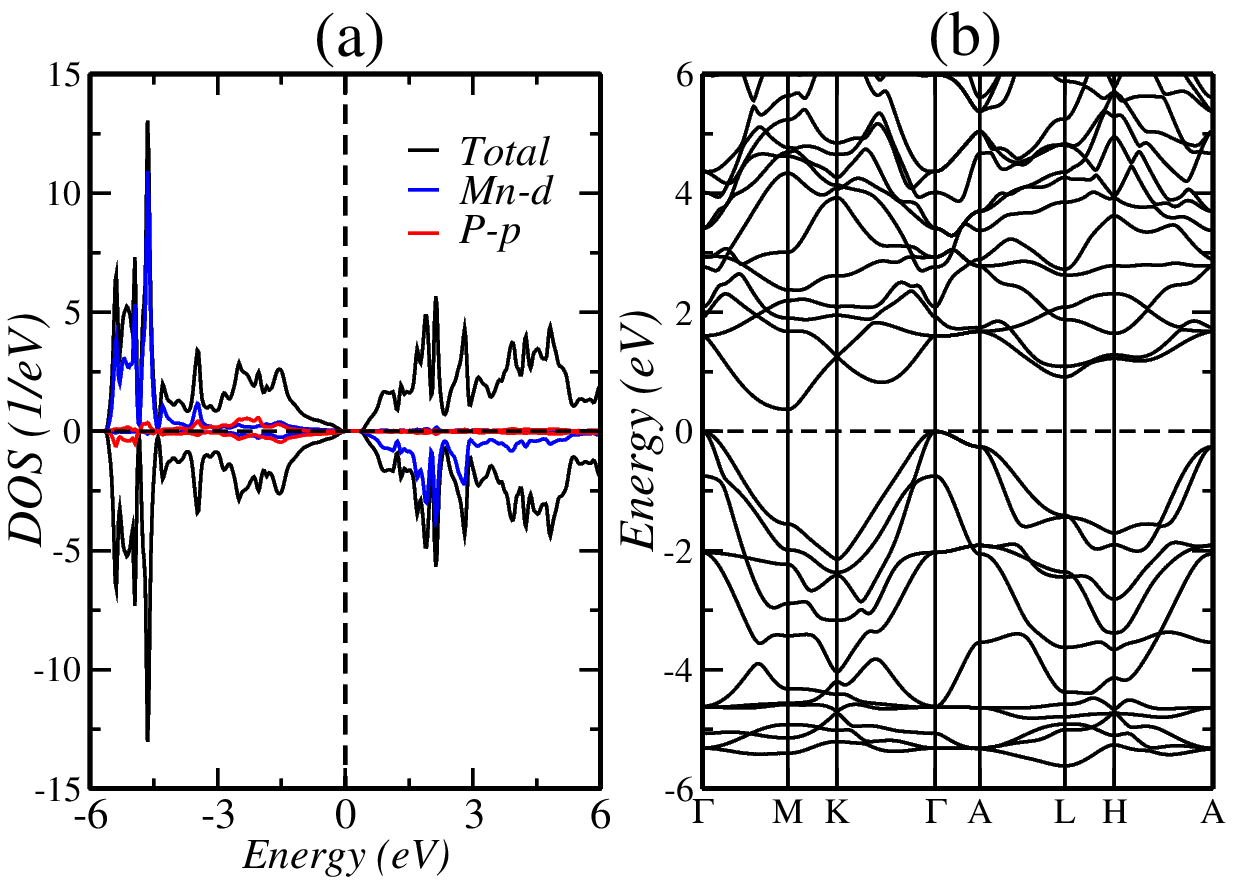}
\caption{ (a) The total density of states and partial density of states of Mn-d and P-p in CaMn$_2$P$_2$ for the AFM state, computed using GGA+U formalism. The
vertical dashed line (Black color) indicates the Fermi energy ($E_F$). (b) Band dispersion along various high symmetry directions of the Brillouin zone, demonstrating the indirect nature of the band-gap in this material.} 
\label{fig2}
\end{figure}

 
The computed total density of states (DOS), partial DOS corresponding to Mn-$d$, P-$p$ and band-dispersion along the high-symmetry directions for the AFM configuration of CaMn$_2$P$_2$ is displayed in Fig.~\ref{fig2}.  The total DOS and band-dispersion reveal that it is an indirect band-gap semiconductor with a gap of 0.4 eV which agrees well with the experimental results~\cite{Zheng2023}. The PDOS of Mn-$d$ reveal that the majority Mn-$d$ states are fully occupied and lie well below the Fermi level, while the minority Mn-$d$ states appear above it. This is consistent with the nominal Mn$^{2+}$ (3$d^5$ occupancy) charge state and a high-spin ($S=\frac{5}{2}$) configuration in accordance with Hund's rule. Thus the spin moment of Mn arises from the five unpaired spin, giving rise to a significantly high value of 4.28 $\mu_B$. A weak hybridization between Mn-$d$ and P-$p$ orbitals is evident from small spectral weight near the Fermi level, while the dominant spectral weight of Mn-$d$ and P-$p$ states lies deeper below. Such electronic structure emphasizes the highly localized nature of Mn-$d$ state and hence a pure Heisenberg like spin Hamiltonian could be an appropriate model to describe the magnetism of this system. 
\par 
\noindent\textbf{Inter-site exchange couplings and Heiseberg Spin-Hamiltonian:} Next, to characterize the magnetic ground state in this system, we computed interatomic exchange interactions ($J_{ij}$s) between neighboring Mn spins using magnetic force theorem \cite{magnetic-force1,magnetic-force2}. The relevant $J_{ij}$s are labeled in Fig.\ref{fig1}(b). Here the 1$^{st}$ nearest neighbor (NN) Mn-Mn distance is 2.92 $\AA$ and it couples Mn spin across layers along the $c$-axis ($J_1$). The 2$^{nd}$ NN (distance = 4.10 \AA) couples Mn within the $a$-$b$ plane ($J_2$). The 3$^{rd}$ NN distance is bit larger (5.03 \AA) and it is also an inter-layer coupling ($J_3$). The calculated exchange coupling constants for these three neighbors are displayed in Table~\ref{Table1}. Our results find that all the couplings are antiferromagnetic in nature. The magnitude of $J_1$ is found to be strongest, $J_2$ is also significantly large (40\% of $J_1$) and $J_3$ is weakest among these couplings (4\% of $J_1$). In this framework, the $J_1$ and $J_3$ promotes long-range antiferromagnetic order, as they are non-frustrated. However, the presence of a sizable $J_2$ introduces magnetic frustration due to the triangular geometry. We note here that in many triangular and honeycomb-based spin systems, a frustrated interaction competes with the non-frustrated one, leading to nontrivial magnetic ground states, including spin liquids and spiral magnetic structures~\cite{PhysRevResearch.2.033260, PhysRevB.77.075119, PhysRevLett.128.227201}. Furthermore, the spin-orbit coupling effects were examined via GGA+U+SOC calculations. Our calculations confirms that antisymmetric Dzyaloshinskii–Moriya interactions, symmetric anisotropic exchanges, and single-ion anisotropy terms are negligible. This result can be directly attributed to the Mn$^{2+}$ ions adopting a high-spin nominal 3$d^5$ configuration, where the orbital angular momentum is strongly quenched by the symmetric filling of the $d$ orbitals. The high-spin state of Mn$^{2+}$ ions is already been discussed from our computed electronic structure (see Fig.~\ref{fig2}). As a consequence, the effective SOC associated with Mn sites is extremely weak, suppressing any significant anisotropic magnetic interactions. Thus, based on the insights obtained from our first-principle calculations, we propose the following isotropic Heisenberg spin model to describe the magnetism of CaMn$_2$P$_2$. 
\begin{equation}\label{eq2}
\ H =  -J_{1} \sum_{<ij>}(\hat{S}_i \cdot \hat{S}_j) - J_{2} \sum_{<ik>}(\hat{S}_i \cdot \hat{S}_k) - J_{3}\sum_{<il>}(\hat{S}_i \cdot \hat{S}_l)
\end{equation}
We solved the above Hamiltonian by employing linear spin wave theory as implemented in spinW code~\cite{Toth_2015}. This approach use an energy minimization procedure, where an initial spin configuration is relaxed using classical numerical optimization techniques. Here we employed the Luttinger-Tisza method~\cite{PhysRev.70.954} to explore potential long-range magnetic orders by identifying wave vectors that minimize the Hamiltonian in reciprocal space.  Our results find that a spin-spiral magnetic state with propagation vector ($\frac {1}{6},\frac {1}{6}, 0$) comes out to be lowest in energy. This is in agreement with the recently reported magnetic structure determined from neutron diffraction experiments on single crystals~\cite{PhysRevB.107.054425}. Thus, it is clear that the minimal spin-model to describe the magnetism of CaMn$_2$P$_2$ is a $J_1$ -$J_2$- $J_3$ Heisenberg model as given in above equation. However, both $J_1$ and $J_3$ couples Mn spins across adjacent layers along the $c$-axis and $J_3$ is almost negligible compared to $J_1$ ((4\% of $J_1$)). We find that $J_3$ does not play a significant role in determining the magnetic ground state. Hence, the essential physics of this system can be captured within a $J_1$ -$J_2$ Heisenberg model, where $J_1$ is a strong, non-frustrated antiferromagnetic (AFM) interaction, while $J_2$ introduces geometrical frustration. The competition between these two interactions is the primary driver of the observed spin-spiral ground state. 
\begin{table}
\renewcommand{\arraystretch}{1.8}
\setlength{\tabcolsep}{5pt} 
\centering
\caption{Calculated Mn–Mn exchange coupling parameters \( J_i \) obtained from GGA+\( U \) calculations. Negative values of \( J_i \) indicate antiferromagnetic interactions. The exchanges \( J_1 \), \( J_2 \), and \( J_3 \), corresponding to distinct Mn–Mn distances, are illustrated in Fig.~\ref{fig1}(b). For each coupling, the number of neighbors \( z_i \), Mn–Mn separation, and the values of \( J_i \) (in meV) are listed. To highlight the relative importance of each interaction, the normalized quantities \( J_i / J_1 \) and \( z_i J_i / J_1 \) are also included.}
\begin{tabular}{c c c c c c}
\hline\hline
Exchange & Number of & Mn-Mn & Values & $\frac{J_i}{J_1}$ & $\frac{zJ_i}{J_1}$ \\
($J_i$) & neighbors ($z_i$) &  distance (\AA) & (meV) &   &  \\ 
\hline
$J_1$ & 3 & 2.92 & $-39.96$ & 1 & 1 \\
$J_2$ & 6 & 4.10 & $-16.02$ & 0.40 & 0.80 \\
$J_3$ & 3 & 5.03 & $-1.72$ & 0.04 & 0.04 \\
\hline\hline
\end{tabular}
\label{Table1}
\end{table}
\par 
We note here that $J_1$ -$J_2$ model has been extensively studied in two dimensional (2D) square-lattice antiferromagnets~\cite{PhysRevB.38.9335,PhysRevB.109.184410}, where ferromagnetic $J_1$ and antiferromagnetic $J_2$ induce frustration, leading to exotic spin textures and quantum disordered states. This model has been experimentally realized in materials such as Li$_2$VOSiO$_4$, Li$_2$VOGeO$_4$, BaCdVO(PO$_4$)$_2$ and SrZnVO(PO$_4$)$_2$ etc where frustration suppresses conventional long-range order and also exhibit spin-liquid ground states \cite{PhysRevLett.85.1318, PhysRevB.78.064422, PhysRevB.81.174424}. In CaMn$_2$P$_2$, we propose a new realization of the $J_1$ -$J_2$ model, where both interactions are antiferromagnetic but act in different planes. 
The $J_1$ governs the coupling across the layers along the $c$-axis (interlayer interaction), while $J_2$ couples Mn spins within the hexagonal $a$-$b$ plane, introducing frustration. Thus CaMn$_2$P$_2$ form a frustrated network of interactions in a true three-dimensional (3D) setting. This geometric distinction places CaMn$_2$P$_2$ in a different universality class compared to the extensively studied 2D $J_1$-$J_2$ square-lattice models~\cite{PhysRevB.38.9335,PhysRevB.109.184410}. 
\begin{figure}[t]
\includegraphics[width=1.0\columnwidth]{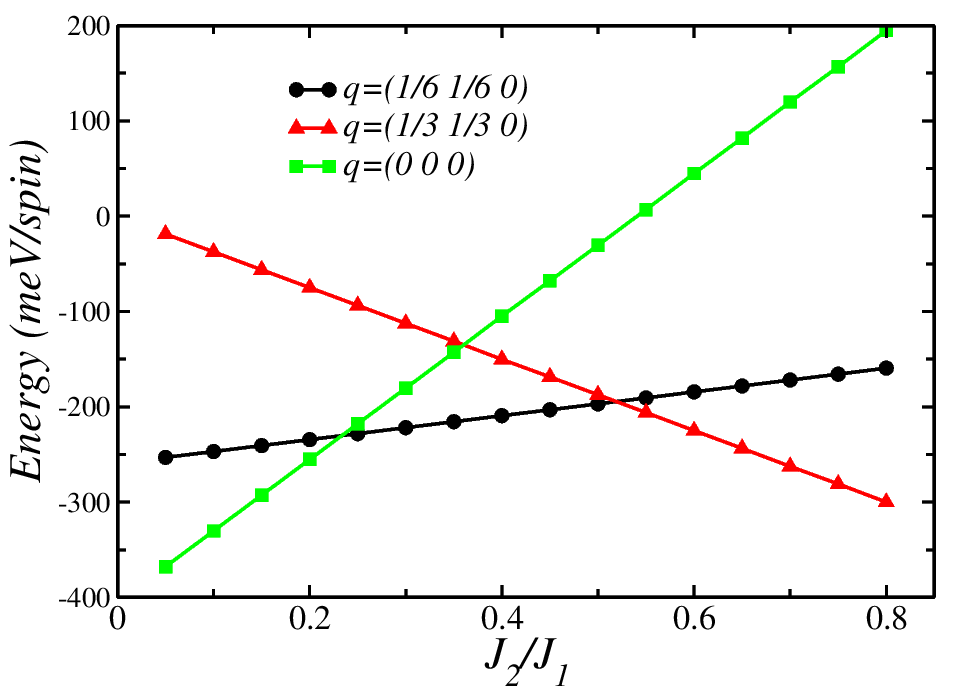}
\caption{ Energies per spin as a function of $\frac {J_2}{J_1}$ ratio for magnetic ground states corresponding to three different ordering vector ($q$-vector) as obtained from linear spin wave theory calculations.} 
\label{fig3}
\end{figure}

\begin{figure*}
 \begin{center}
\includegraphics[width=2\columnwidth]{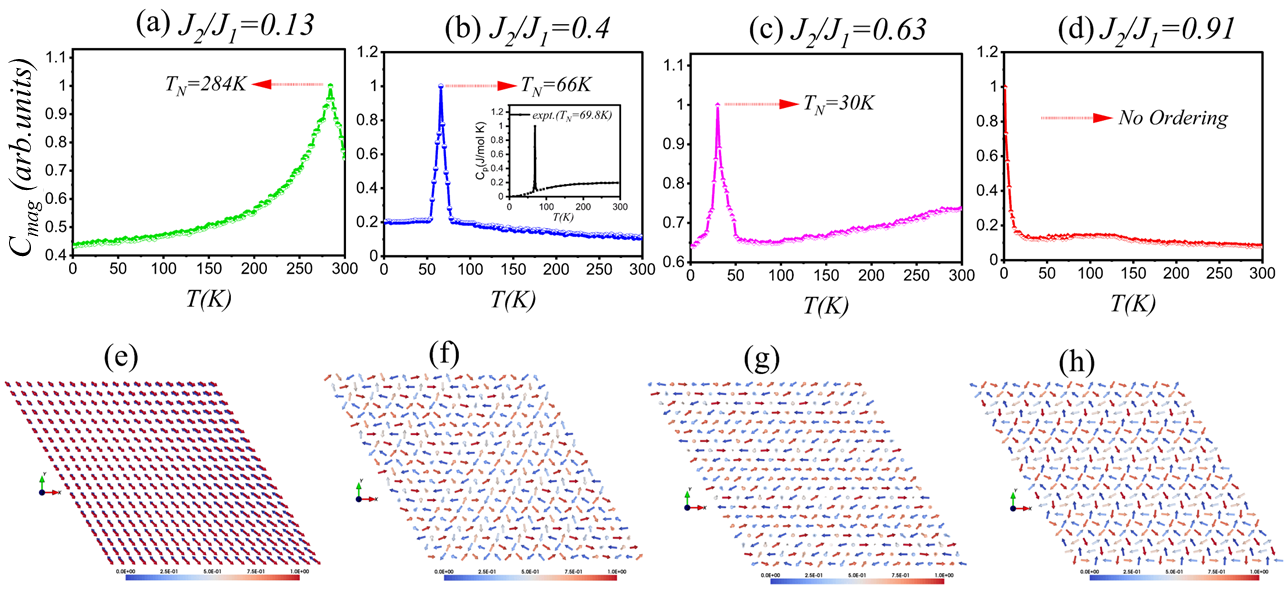}
   \caption{Temperature dependent magnetic specific heat ($C_{mag}$) as calculated using atomistic spin dynamics simulations for $\frac{J_2}{J_1}$ = (a) 0.13, (b) 0.4, (c) 0.63, and (d) 0.91. The obtained ordering temperatures ($T_N$) from the peaks of $C_{mag}$ are displayed. The inset of fig. (b) shows the experimental data of ref.~\cite{doi:10.1073/pnas.2108724118}. The visualization of spin configurations in the ground state are shown in lower pannel in (e) to (h) for each $\frac{J_2}{J_1}$, respectively.}
 \label{fig4}
 \end{center}
\end{figure*}

\par 
\noindent\textbf{Magnetic ground state using linear spin wave theory:}
Since our results demonstrate that a $J_1$ -$J_2$ model provides a minimal yet robust theoretical framework for understanding the magnetism in CaMn$_2$P$_2$, we explored the possibilities of different exotic magnetic ground states arising due to the delicate balance between these two competing interactions. In practice, the various magnitude of $J_1$ and $J_2$ could be achieved by chemical substitution, pressure, or applying strain in this class of materials having CaAl$_2$Si$_2$-type crystal structure. We carried out linear spin wave theory calculations as implemented in spinW code~\cite{Toth_2015} for a few $\frac {J_2}{J_1}$ ratio, keeping the nature of coupling (i.e. antiferromagnetic) intact. We kept the magnitude of $J_1$ fixed and tuned $J_2$'s magnitude for changing $\frac {J_2}{J_1}$ ratio. The obtained spin-spiral $q$-vector as a function of $\frac{J_2}{J_1}$ ratio is shown in fig.~\ref{fig3}. Our calculations correctly predicts the magnetic propagation $q$-vector = ($\frac {1}{6},\frac {1}{6}, 0$) in the range  $0.23 < \frac {J_2}{J_1} < 0.52$. For CaMn$_2$P$_2$, the obtained $\frac {J_2}{J_1}=0.4$  and thus it fall within this range. Further, our calculation predicts that by lowering this ratio, system could be driven to the N\'eel antiferromagnet state ($q$ = (0, 0, 0)) for $\frac {J_2}{J_1}$ less than 0.23. Similar ordered N\'eel's state in bipartite honeycomb layer structure has also been reported in case of dominating nearest neighbour interaction~\cite{PhysRevB.94.184422,PhysRevB.96.064413}. On the other hand, for $\frac {J_2}{J_1}$ ratio greater than 0.52, magnetic ground state transit to $q = (\frac {1}{3},\frac {1}{3}, 0)$ spin-spiral state. Our findings reveal that magnetic systems characterized by two dominant exchange interactions, a frustrated next-nearest-neighbor interaction ($J_2$) and a non-frustrated nearest-neighbor exchange ($J_1$) exhibit nontrivial magnetic ground states arising from their competing nature. The interplay between these interactions can lead to various emergent magnetic phases, including spin-spiral states, N\'eel order, depending on the relative strengths of $J_1$ and $J_2$. This model provides a fundamental framework for understanding the magnetic properties of a broader class of compounds with layered honeycomb structures (CaAl$_2$Si$_2$-type), where geometric frustration and competing interactions play a crucial role in dictating the ground state spin configurations. \\
\noindent\textbf{Magnetic transitions using atomistic spin dynamics simulations:}
In order to give further credence to the above conclusion and also to find out the magnetic transition temperatures corresponding to different magnetic states, we performed atomistic spin dynamics simulations by solving the above spin Hamiltonian (Eq.~\ref{eq2} ) using the stochastic Landau-Lifshitz-Gilbert (sLLG) equation as implemented in UppASD~\cite{PhysRevB.54.1019,Skubic_2008} code. The calculated temperature dependent magnetic specific heat ($C_{mag}(T)$) for the estimated ratio of $\frac{J_2}{J_1}$ in CaMn$_2$P$_2$ is displayed in Fig.~\ref{fig4}(b). The sharp peak in $C_{mag}(T)$ shows that the ordering temperature ($T_N$) comes out to be 66 K which is in good agreement with experiment~\cite{doi:10.1073/pnas.2108724118,PhysRevB.107.054425} (69.8 K). Further the overall agreement between the theoretical and experimental  $C_{mag}(T)$ (see inset of Fig.~\ref{fig4}(b)) demonstrate the accuracy of our computed exchange parameters that are used in constructing the spin Hamiltonian for this system. The spin-texture is calculated and the orientation of Mn spin moments in the hexagonal $a-b$ plane (Fig.~\ref{fig4}(f)) show a spin-spiral magnetic ground state with ordering vector $q = (\frac{1}{6}, \frac{1}{6}, 0)$. Thus, we are able to explain the magnetism of CaMn$_2$P$_2$ with a minimal isotropic Heisenberg spin model, comprising primarily 1$^{st}$ and 2$^{nd}$ NN couplings. As expected, by tuning the $\frac{J_2}{J_1}$, we are not only able to alter the magnetic ground states, but also could tune the ordering temperatures as shown from the computed $C_{mag}(T)$ in Fig.~\ref{fig4}(a)-(c). The ordering temperature for the antiferromagnetic N\'eel state is very high (284 K) for the chosen value of magnetic couplings, while the spin spiral transition with $q = (\frac{1}{3}, \frac{1}{3}, 0)$ occurs at relatively lower temperature (30 K). The computed spin-texture for both these cases are displayed in Fig.~\ref{fig4}(e) and (g) which correctly show the ordering vectors. Beyond a critical value of $\frac{J_2}{J_1} > 0.91$, we don't find any signature of magnetic ordering as evident from the $C_{mag}(T)$ plot in Fig.~\ref{fig4}(d). The absence of any long range ordering and signature of disordered phase is also  observed in the computed spin texture shown in Fig.~\ref{fig4}(h). This could be attributed to the dominance of strong magnetic frustration, giving rise to the emergence of complex magnetism as T approaches 0 K. 
\par
\begin{figure*}[t]  
   \begin{center}
       \includegraphics[width=1\linewidth]{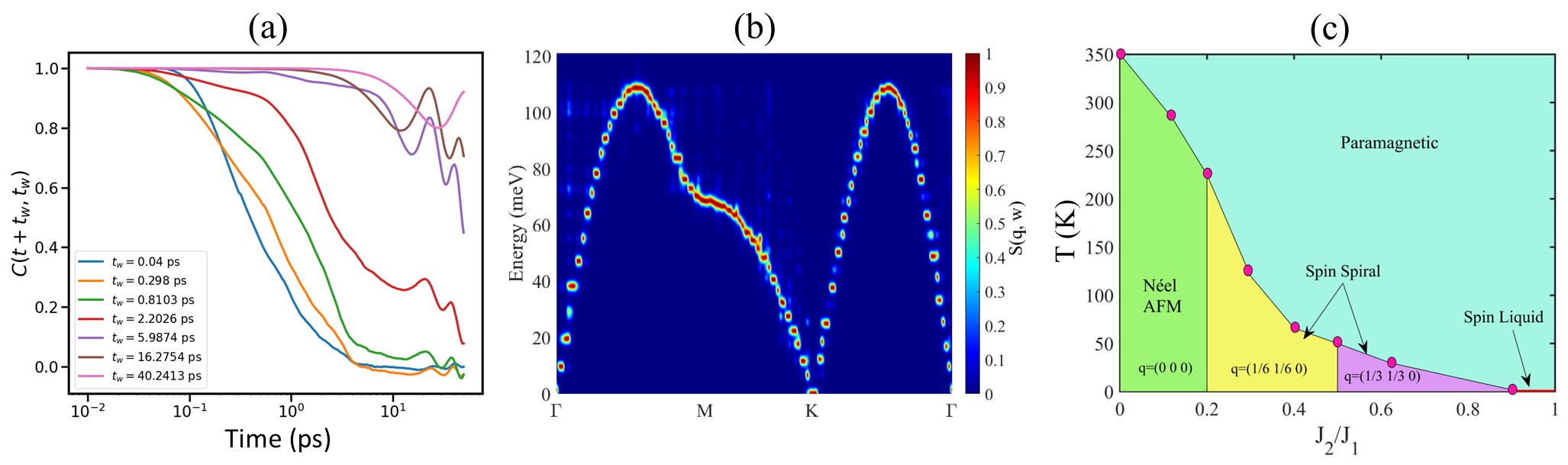} 
    \caption{(a) The time dependent spin autocorrelation function $C(t+t_w,t_w)$ for different waiting times ($t_w$) . (b) Dynamical structure factor $S(\mathbf{q}, w)$ computed along high-symmetry directions of the Brillouin zone, illustrating the momentum and energy-resolved spectrum. (c) The magnetic ground states for different range of $\frac{J_2}{J_1}$ and the corresponding ordering temperature are schematically presented in this phase diagram. The red region for $\frac{J_2}{J_1} > 0.91$ is the spin liquid state where $T_N$ approaches 0 K. }
    \label{fig5}
    \end{center}
\end{figure*}

 
In order to understand this interesting part of the phase diagram corresponding to high value of $\frac{J_2}{J_1}$ and very low temperature, we studied the relaxation properties of system by computing the time-displaced spin-spin autocorrelation function as given below:
\begin{equation}
   \label{eq:ac}
   C_0(t+t_w,t_w) = \langle \overrightarrow{S_i}(t+t_w)\cdot\overrightarrow{S_i}(t_w) \rangle ,
\end{equation}
where waiting times, $t_w$ represents the time duration for which the system is allowed to evolve after initialization but before the measurement of spin correlations begins. This facilitates the analysis of spin correlations, offering insight into the underlying dynamical behavior of the system. In an ergodic system that has reached equilibrium, the autocorrelation function becomes invariant under time translations and depends only on the time difference $t$, rendering $t_w$ irrelevant. However, in systems with frustration or disorder, the dynamics can become markedly slow, preventing the system from equilibrating over the accessible time scales. In such a non-ergodic scenario, the autocorrelation function retains an explicit dependence on $t_w$, reflecting the system’s memory of its initial conditions. Therefore, the explicit dependence of $C_0(t+t_w,t_w)$ on the waiting time $t_w$ acts as a sensitive indicator of aging effects and slow relaxation, offering valuable insights into the underlying magnetization dynamics within the frustrated region of the phase diagram. The computed $C_0(t+t_w,t_w)$ at logarithmically spaced $t_w$ is shown in Fig.~\ref{fig5}(a). For all $t_w$, the autocorrelation initially starts at unity. For the short waiting times (upto $t_w$=0.8103 ps), the correlation function exhibits a rapid decay and become 0.0 within 10 ps, indicating fast relaxation. As $t_w$ increases, the decay becomes progressively slower, with long-time correlations persisting. At the largest $t_w$ values, we observe presence of oscillations and non-monotonic behavior which suggests aging effects. Importantly, this observed non-ergodic behavior arises intrinsically from the sLLG spin dynamics governed by the underlying spin Hamiltonian used in our simulations. Therefore, the persistent fluctuations and slow relaxation dynamics at large waiting times are consistent with the highly frustrated nature of the system, possibly exhibiting spin-liquid-like behavior. The overall nature of the autocorrelation, in particular the appearance of plateau at longer times indicate that the system possess dynamical correlation and hence it is a potential spin liquid candidate. Thus, the data in Fig.\ref{fig5}(a). suggest a possible spin liquid phase as T approaches 0 K and this occurs when the magnetic frustration is large. 
\par 
To further probe the nature of the highly frustrated magnetic phase, we have computed the dynamical spin structure factor, $S(\mathbf{q}, w)$, at zero temperature for a representative parameter regime characterized by a large ratio $\frac{J_2}{J_1}$, where $J_1$ = -39.96 meV , and $J_2$ = -36.50 meV. The quantity $S(\mathbf{q}, w)$ captures spin excitations resolved in momentum and energy, and is defined as the space-time Fourier transform of the spin-spin correlation function, expressed classically as:
\begin{align}
S(\mathbf{q}, w) &= \sum_{\alpha\beta} \left( \delta_{\alpha\beta} - \frac{q_\alpha q_\beta}{q^2} \right) 
\frac{1}{2\pi N} \sum_{\mathbf{i,j}} e^{i \mathbf{q} \cdot (\mathbf{r_i}-\mathbf{r_j})} \nonumber \\
&\quad \times \int_{-\infty}^{\infty} e^{i \omega t} \langle S_i^\alpha(t) S_j^\beta(0) \rangle\ dt,
\label{Sqw}
\end{align}
where, $\alpha$ and $\beta$ denote the cartesian components of the spins ($S_i$, and $S_j$), $q$ is the momentum, $w$ is energy, and and $N$ is the number of lattice sites. Figure~\ref{fig5}(b) displays the calculated $S(\mathbf{q}, w)$ along the high-symmetry path $\Gamma-M-K-\Gamma$ in the Brillouin zone. The spectrum exhibits broad, dispersive features extending up to approximately 110 meV. Notably, the spectral intensity is highly anisotropic across momentum space. Strong enhancements in spectral weight are observed around the $M$ point and in the intermediate region between $\Gamma$ and $M$ as well as between $\Gamma$ and $K$ points, particularly at higher energies, whereas the other parts displays a relatively suppressed intensity. This non-uniform distribution of spectral weight reflects the impact of strong magnetic frustration and the absence of sharp, coherent magnon modes typically associated with long-range magnetic order. The broad, incoherent nature of the excitations and the momentum-selective intensity enhancements are indicative of a magnetically disordered ground state. This is consistent with a scenario in which dominant antiferromagnetic $J_2$ interactions on the triangular sublattice destabilize long-range order, pushing the system toward a dynamically fluctuating phase, possibly proximate to a spin liquid ground state. The observed features in our calculations bear close resemblance to those reported in prior theoretical studies of frustrated triangular-lattice Heisenberg models~\cite{PhysRevB.88.094407,PhysRevX.9.031026}. These findings are also in agreement with our analysis of the spin-spin autocorrelation function, reinforcing the picture of a dynamically disordered ground state governed by geometric frustration and  competing interaction.
\par 
Thus, we explore the magnetism for different $\frac{J_2}{J_1}$ ratio and found an interesting phase diagram as shown in Fig.~\ref{fig5}(c) . 
The phase diagram demonstrates the crucial role of competing interactions in stabilizing diverse magnetic phases. For small $\frac{J_2}{J_1}$, the system favors a collinear antiferromagnetic (AFM) order with wavevector $q$=(0,0,0), dictated by the dominant unfrustrated nearest-neighbor interaction $J_1$. As 
$\frac{J_2}{J_1}$ increases, frustration leads to the emergence of incommensurate spin spiral phases with wave-vectors $q=(\frac{1}{6},\frac{1}{6},0)$ and $q=(\frac{1}{3},\frac{1}{3},0)$, where spins adopt a non-collinear arrangement as a compromise between competing exchange interactions. When $\frac{J_2}{J_1}$ approaches unity, frustration becomes dominant, completely destabilizing long-range magnetic order and giving rise to a spin liquid like phase, characterized by persistent local spin correlations and strong fluctuations. The frustration-induced phase transitions have been widely observed in geometrically frustrated systems ~\cite{frustmag}, including triangular, kagome, and pyrochlore lattices as well as in materials such as YbMgGaO$_4$ and SrCu$_2$(BO$_3$)$_2$, where similar competition between exchange couplings drives exotic magnetic ground states \cite{PhysRevLett.85.1318, PhysRevB.78.064422, PhysRevB.81.174424}. The computed transition temperature ($T_N$) decreases with increasing frustration, as long-range order is progressively suppressed, leading to a paramagnetic phase at high temperatures. This phase diagram elegantly demonstrates how competing nearest and next-nearest neighbor interactions can generate a rich variety of magnetic phases. More broadly, our findings indicate that modifications in exchange interactions (e.g., via chemical substitution or external pressure) can drive significant changes in magnetic order. This explains why similar material like CaMn$_2$Bi$_2$~\cite{PhysRevB.91.085128} exhibit N\'eel antiferromagnetic ($q$ = (0,0,0)) ground states despite their shared structural framework. The phase diagram derived from our study serves as a unifying model for predicting and explaining magnetic behavior in a broad range of trigonal 122-type Mn-based compounds. We note here that the value of $T_N$ in this phase diagram (Fig.~\ref{fig5}(c)) depends on the magnitudes of $J_1$ and $J_2$. However, the appearance of different magnetic ground states is robust with respect to the $\frac{J_2}{J_1}$ ratio.
\section{Conclusion}
In conclusion, our detail first-principles calculations and spin-dynamics simulations, have demonstrated that the spiral magnetic ground state of CaMn$_2$P$_2$ can be described using an isotropic Heisenberg spin model, consisting upto 3$^{rd}$ NN exchange interactions. However, the predominant couplings are antiferromagnetic nearest-neighbor interaction $J_1$ between the two Mn-plane along crystallographic $c$-direction and a frustrated next-nearest-neighbor interaction $J_2$ in the $a$-$b$ plane where Mn forms a triangular arrangement. We show that this interplay between competing interactions is responsible for stabilizing a spin-spiral ground state with $q$=($\frac{1}{6},\frac{1}{6},0$), in excellent agreement with experimental neutron diffraction data~\cite{PhysRevB.107.054425}. Our computed magnetic specific heat also nicely agrees with experiment and provide a good estimate of the observed transition temperature~\cite{PhysRevB.107.054425}. These results conclude that an isotropic $J_1$-$J_2$ Heisenberg model is able to describe the magnetism of this system and also confirm the accuracy of our estimated $J_{ij}$s from first principles calculations. 
\par 
Further, by varying the $\frac{J_2}{J_1}$ ratio, we establish a comprehensive phase diagram, revealing transitions from N\'eel order to different spin-spiral states and, eventually, to a highly frustrated magnetic phase. Notably, when  $\frac{J_2}{J_1}$ approaches unity, long-range order is suppressed, and the system exhibits signatures of a spin-liquid-like state, characterized by slow relaxation dynamics and persistent spin fluctuations. Hence, our results conclude that CaMn$_2$P$_2$ and broadly this class of compounds with layered honeycomb structures (CaAl$_2$Si$_2$-type) is a realization of $J_1$-$J_2$ model in 3D lattice. Thus, we presents a fundamentally different geometric realization of the  $J_1$-$J_2$ model compared to the well studied 2D $J_1$-$J_2$ model where both interactions lie on the same-plane. Despite these significant structural differences, our calculations reveal a similarly rich phase diagram, including N\'eel order, spin-spiral states, and a possible spin-liquid-like phase at large frustration, demonstrating that the interplay of competing exchange interactions in 3D systems can yield phenomena analogous to their lower-dimensional counterparts. Thus, our details analyses demonstrate the ability to tune the ground state through frustration and offers a pathway for exploring novel magnetic phases in similar three-dimensional frustrated systems.
\section{ACKNOWLEDGMENTS}
S.K.P acknowledges support from DST SERB grant (CRG/2023/003063). The computations were enabled by resources provided by Bennett University.

\end{document}